# *Caenorhabditis elegans* and the network control framework – FAQs

Emma K. Towlson[1], Petra E. Vértes[2], Gang Yan[1,3], Yee Lian Chew[4], Denise S. Walker[4], William R. Schafer[4], Albert-László Barabási[1,5,6,7*]

[1]*Center for Complex Network Research and Department of Physics, Northeastern University, Boston, Massachusetts, 02115, USA*
[2]*Department of Psychiatry, Behavioural and Clinical Neuroscience Institute, University of Cambridge, Cambridge, CB2 0SZ, UK*
[3]*School of Physics Science and Engineering, Tongji University, Shanghai 200092, P. R. China*
[4]*Division of Neurobiology, MRC Laboratory of Molecular Biology, Cambridge Biomedical Campus, Francis Crick Avenue, Cambridge, CB2 0QH, UK*
[5]*Center for Cancer Systems Biology, Dana Farber Cancer Institute, Boston, Massachusetts, 02115, USA*
[6]*Department of Medicine, Brigham and Women's Hospital, Harvard Medical School, Boston, Massachusetts, 02115, USA*
[7]*Center for Network Science, Central European University, H-1051, Budapest, Hungary*

*Corresponding author; email alb@neu.edu

**Abstract**

Control is essential to the functioning of any neural system. Indeed, under healthy conditions the brain must be able to continuously maintain a tight functional control between the system's inputs and outputs. One may therefore hypothesise that the brain's wiring is predetermined by the need to maintain control across multiple scales, maintaining the stability of key internal variables, and producing behaviour in response to environmental cues. Recent advances in network control have offered a powerful mathematical framework to explore the structure-function relationship in complex biological, social, and technological networks (1–3), and are beginning to yield important and precise insights for neuronal systems (4–12). The network control paradigm promises a predictive, quantitative framework to unite the distinct datasets necessary to fully describe a nervous system, and provide mechanistic explanations for the observed structure and function relationships.

Here, we provide a thorough review of the network control framework as applied to *C. elegans* (4), in the style of a FAQ. We present the theoretical, computational, and experimental aspects of network control, and discuss its current capabilities and limitations, together with the next likely advances and improvements. We further present the Python code to enable exploration of control principles in a manner specific to this prototypical organism.

**Introduction**

Connectomics has entered an era of rapid advances on an industrial scale (13–17) which will, in the next few years, offer datasets of unprecedented size and exquisite detail pertaining to the brain's wiring diagram. The theoretical and computational challenges that accompany these advances are also unprecedented: how to handle the unwieldly amount of data, to incorporate and tie together diverse information types such as precise neuronal morphologies and genetic profiles, and how to build experimentally tractable hypotheses and predictions. New tools must be developed in order to tackle this enormous challenge, and any such tools – designed to handle diverse data types in the context of one system – will almost certainly have to cross the traditional disciplinary borders.

Network control has been showing potential as one such tool (3,6,18). Control in the context of the brain may be thought of in two very distinct ways. Firstly, and perhaps most intuitively, control could be used to design a perturbation in order to drive certain brain functions to a desired state (12,19,20). But secondly, control may be also thought of in the sense of understanding how the brain *itself* controls behaviour under normal conditions, and elucidating the structural requirements to facilitate this. Studying control in this second way can reveal fundamental organising principles and mechanisms pertaining to the function of neuronal systems, also constraining their connectome.

To demonstrate the importance of control principles in neural systems, we recently framed the locomotion response in *C. elegans* as a target control problem (4). In doing so, we simultaneously provided the first falsifiable experimental proof of the utility of network control principles in a real system and recovered new insights about previously unknown neuron function in locomotion. In this paper, we aim to make the control-based approach above accessible to a wider and interdisciplinary audience, in terms of understanding, ability to implement, and inspiration to develop. To this end, we adopt a slightly unconventional style, following a 'Frequently Asked Questions – or 'FAQ'- format. The 'Q's are grouped to discuss: (i) The theoretical framework, or *Network control framework*; (ii) The *Model assumptions* underlying the application to *C. elegans*; (iii) the *Computational and experimental details,* including a link to Python code with which to implement the analysis; (iv) *Potential improvements* to the current framework; (v) Generalisation to further behaviours and organisms and other *Future perspectives*.

*NETWORK CONTROL FRAMEWORK*

**How does a target control framework apply to *C. elegans* locomotion?**

*C. elegans* moves in a sinusoidal fashion, via dorso-ventral bends shaped by its 95 rhomboid body wall muscle cells. Mechanosensory cues (like gentle touch) elicit a locomotory response, and we found that such behaviours map very naturally onto a target control problem as follows: the behaviour is driven by stimuli to sensory neurons (control signals to input nodes),

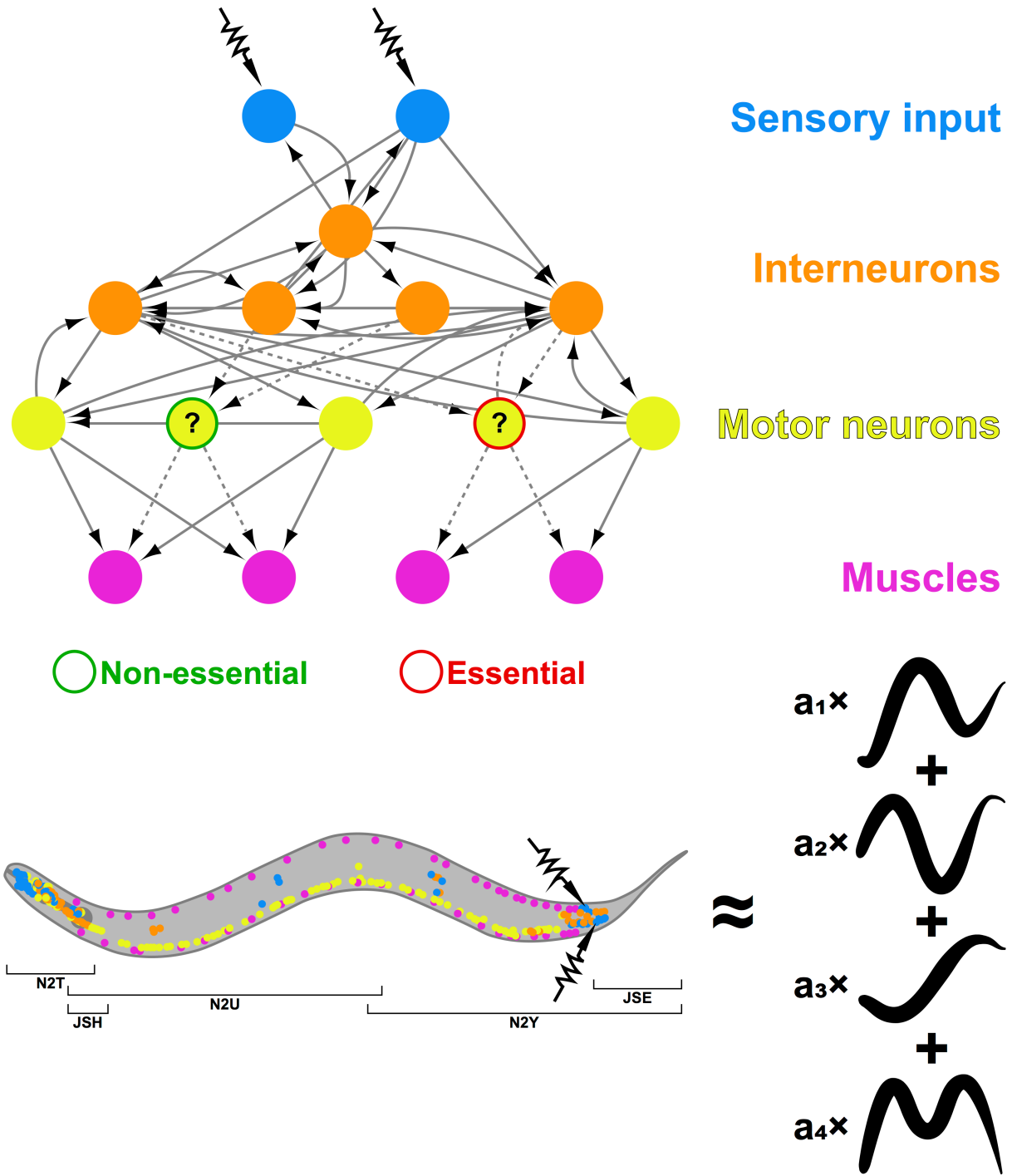

**Figure 1: The network control approach to understanding the behavioural responses of *C. elegans*.** Graphical representation of the proposed control framework, adapted from (4). According to structural controllability in the context of a locomotory response to stimuli, if removal of a neuron disrupts controllability of the muscles, we designate it "Essential" for locomotion; if not, we call it "Non-essential". To make this assessment, we first mapped the *C. elegans* responsive locomotor behaviours into a target network control problem, asking to what degree the sensory neurons (blue) can control the muscles (pink). This allowed us to predict the previously-unknown involvement of PDB in *C. elegans*

locomotion, and functional differences between individual neurons within the DD neuronal class. We test our predictions through cell-specific laser ablation and worm tracking experiments, and statistically comparing eigenworm features. The original EM images in (21) were reconstructed from five partial worms – primarily N2U and JSE (adult hermaphrodites), then N2T for the anterior nerve ring (adult hermaphrodite), N2Y (adult male) for the section between N2U and JSE, and finally JSH (L4 larva) to check connectivity in the nerve ring. Adapted from (64) and (4).

*C. elegans* moves in a sinusoidal fashion, via dorso-ventral bends. Its 95 rhomboid body wall muscle cells are arranged as staggered paired rows in four quadrants (dorsal left/right and ventral left/right), and each muscle cell receives multiple inputs from some of the 75 motor neurons. Corresponding muscles contract and relax in a reciprocal fashion (e.g. for a dorsal bend, the dorsal muscle cells contract while their ventral counterparts relax), and movement requires these waveforms to be propagated sequentially to neighbouring muscle cells, along the length of the animal, in the correct direction. For movement to be sustained, oscillation between the contracted and relaxed states is required. The structure of the motor circuit is critical to achieving these basic requirements. The motor neurons themselves receive input from the "command" interneurons, which constitute a bi-stable circuit that determines the direction of movement, depending on input from sensory neurons (reviewed by (65–67)).

which is then processed by the connectome (control system), and results in the muscle contractions and relaxations that produce locomotion (states of the output nodes) – see Figure 1. In this framework, the intact system, i.e. the connectome as mapped by (21) and subsequently updated by (22,23), informs us of the level of controllability that exists naturally in the worm. We quantify this by the number of linearly independent control signals that reach the muscles. By systematically ablating neurons and neuron classes *in silico*, we can then assess each neuron's impact on controllability by pinpointing the sets of muscles which receive a reduced number of independent signals. This leaves us with a list of neurons that are predicted to play a role in the control of a certain behaviour, predictions that we examine via cell specific laser-ablation and worm tracking experiments.

**How does the control framework compare to other network based predictive tools?**
One can design multiple plausible network topology based approaches (24) to elucidate the neurons important for locomotion. Most of these approaches depend on ranking the neurons based on some network property. The most obvious problem presented by such approaches is the lack of objective criteria for what aspect of a neuron's connectivity should be considered (degree, number of connections to muscle cells, and so on), and lack of criteria to define the cut-off value above which a neuron is deemed essential. Nevertheless, we can clearly learn a lot from such simple networks based approaches (23,25). On the other end of the scale, detailed modelling approaches that take realistic neuronal dynamics into account will ultimately offer vital understanding of the circuits (26–28). The control approach offers a useful middle ground – it does not demand knowledge of detailed dynamics (see below), yet is capable of identifying neurons with important role in control, even if they are not obviously well-connected (e.g. PDB) as well as differences between neurons that on the face of it look similar in their wiring (such as the DDs) (4).

**What do we expect to observe when control theory predicts loss of control for a specific neuron?**

As described above, we can quantify controllability by the number of linearly independent control signals received by the muscles – this also corresponds to the number of muscles which could, in theory, be moved *independently*. Naturally, the worm does not *need* the ability to independently control every single one of its 95 muscles (29), and this is also reflected in our results, which show that fewer than 95 linearly independent control signals can reach the muscles. What the network control framework allows us to do is predict the deviation from the healthy starting point – if the ablation of a neuron reduces the number of control signals reaching the muscles, then we expect a reduction in the worm's ability to finely control locomotion. From this, we can infer if the ablated neuron(s) play a role in the control process. This approach does not tell us the precise role of the ablated neuron – it only tells us that in the absence of the neuron the network loses some degree of control over the muscles. Note also that a loss of controllability does not imply loss of activity: a loss of controllability over a small number of muscles means only that these muscles cannot be independently controlled by the nervous system and must have an amount of correlation. Thus, the predicted phenotypes (reduction in overall controllability of a few muscles) can be in some cases quite subtle.

While the structural controllability framework is deterministic, it allows us to assign a probability to which muscles are affected. This exploits the fact that there are multiple solutions to the control problem (30), each of which give rise to the same level of controllability. Hence cataloguing the independent solutions can inform us which muscles are more likely to experience a reduction in control. In practice, we find that the sets of affected muscles tend to be spatially co-localised across different solutions. For example, in the case of the DD neurons, the control analysis predicted a reduction on control over the set of posterior muscles where defects were experimentally observed (4).

*MODEL ASSUMPTIONS*

**Can we factor in the effect of different connection types on network control?**

Neuronal connections may be: (i) inhibitory or excitatory; (ii) weighted, in a structural or functional sense; (iii) chemical synapses or electrical gap junctions; (iv) synaptic (wired) or extrasynaptic (wireless).

(i) *Inhibitory vs excitatory*. Depending on what kind of ion channel they control, chemical synapses can either be excitatory (by opening sodium/cation channels) or inhibitory (by opening chloride/anion channels). In a network sense, an excitatory synapse is described as a link with a positive sign, and an inhibitory synapse with a negative sign. While the excitatory/inhibitory nature of individual neuromuscular synapses is known, for neuron to neuron synapses it is mostly unknown (in *C. elegans*, all classical neurotransmitters can be either excitatory or inhibitory).

Structural controllability does not make any assumptions about the signs of the links, only whether they are non-zero (2,3). The inhibitory/excitatory nature of the synapses only becomes important if we aim to actually control the network in specific ways, and does not change the conclusions and predictions regarding the more fundamental question of controllability.

(ii) *Weighted*. Similar to the signs of the links, weights are treated as free parameters in the structural controllability calculations. If we are interested in quantities such as control energy, or control time, the weights become essential (2,3,31,32). These weights may be defined structurally (synaptic sizes and numbers), or functionally (correlations between neuronal activity), a choice that should be carefully considered in the context of the line of inquiry.

(iii) *Chemical and electrical synapses,* the two distinct forms of wiring between neurons, differ greatly mechanistically, with effects on connectivity that are felt at the network level – see Figure 2. Specifically, in chemical synapses, an electrical signal in the presynaptic cell is transformed to a chemical signal (release of a neurotransmitter) and then transformed back to an electrical signal in the responding cell (neuron or muscle) through postsynaptic neurotransmitter receptors that are, or control, ion channels. The signal is directional (from presynaptic to postsynaptic cell), and the strength and timing of the signaling depends only on the state (i.e. membrane potential) of the sending cell. Despite some nonlinearities, the properties of chemical synapses can be reasonably approximated by the matrix formulation of our control framework.

In contrast, electrical synapses are channels by which electrical current can flow between coupled cells. While current can in principle flow in either direction through the electrical synapse, at any given time it can only flow in one direction, and this direction is determined by the relative membrane potentials of the coupled cells. Consequently, gap junctions can lead to partial electrical coupling between cells, making their membrane potentials more similar to one another. These connections are less well-modeled by the matrix formulation of our control framework, and they are considerably more restricted than chemical synapses in their ability to transmit a control signal. Moreover, many gap junctions, in *C. elegans* and other organisms, are asymmetric in the expression of their constituent innexins, and consequently pass current more easily in one direction than another. Without knowing which of the 25 innexins is expressed at a particular gap junction (33,34), it is impossible to infer from connectome data which of them are asymmetrically-rectifying, further complicating the inclusion of gap junctions in the control framework.

In our control analysis (4), we treated all synaptic connections as if they were the same, and we used two directed connections (one each way) to represent an undirected gap junction. However, due to the properties noted above, we now suspect that the inclusion of gap junctions in the network may lead to overestimation of the structural controllability of the real connectome – see *Potential improvements* section below.

(iv) *Wired vs wireless*. In addition to synaptic ("wired") connections, neurons also signal to each other using neuromodulators such as monoamines and peptides, connections that are mostly extrasynaptic ("wireless") (35). In other words, neuromodulatory molecules are released by neurons into the system, and this form of signaling is received, on a local and/or global scale, by all neurons which express the relevant receptors. Since the connectome data used for our analysis relied only on the chemical synapses and gap junctions (wired connections), it did not take account of neuromodulatory interactions. In principle, one can describe the wireless connections as external control signals applied to receiving nodes, and, with a sufficiently complete map, one could determine which nodes are required for controllability by neuromodulators. Although we do not yet have such maps, efforts are underway to extend our knowledge of neuromodulatory networks, so in the future this approach may be feasible.

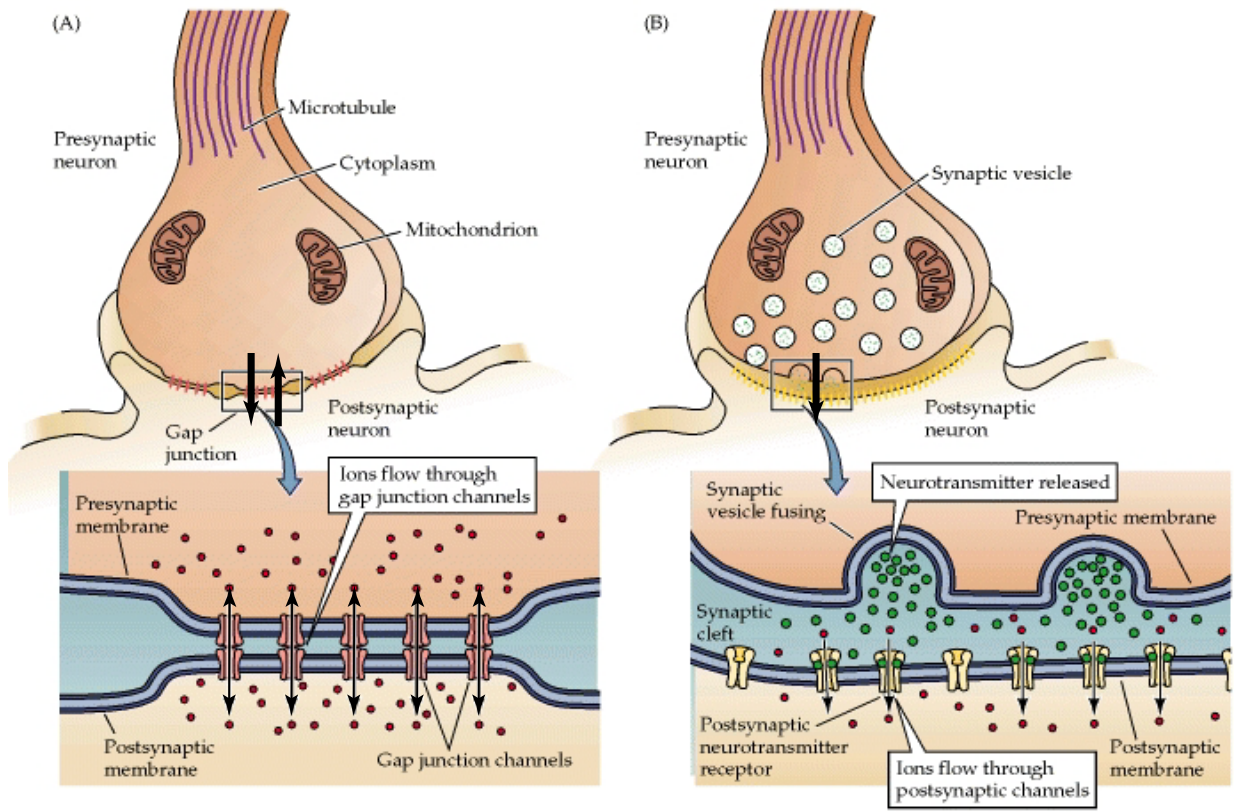

**Figure 2: Chemical synapses and electrical gap junctions.** Chemical synapses and electrical gap junctions have very different properties and underlying mechanisms. In electrical gap junctions (left), voltage is transferred via touching membranes and signals may pass in both directions. In chemical synapses (right), signals are transferred through ion channels from the pre- to post-synaptic neuron.

**Can we use linear dynamics to describe the control principles of a highly nonlinear brain?**

Neuronal dynamics are inherently nonlinear, so if we aim to capture the system's behaviour in detail, we must model it using a fully nonlinear framework, which is currently intractable. Our goal is different here: we aim to understand the role of the network wiring diagram in control. As we discuss next, linear dynamics offer a useful approximation for this purpose in some cases. Recent experimental and numerical studies indicated that the neuronal dynamics of the nematode *C. elegans* are low-dimensional and can be understood as transitions between different attractors (limit cycles or fixed points) (36,37). This allows us to apply *local* controllability, i.e. examine the dynamical equation at the fixed points or along the limit cycles.

If a system is locally controllable along a specific trajectory (such as limit cycles here) in the state space, then the corresponding non-linear system is also controllable along the same trajectory (38). Hence the controllability of the linearised system is expected to illuminate the controllability of neuronal dynamics of *C. elegans* within an attractor, even if some level of uncertainty remains (e.g. if the linear system is not controllable, then the non-linear system may or may not be controllable). Indeed, simulations show that the nonlinear controllability of motifs with non-identical link weights exhibits the same properties as its linear and structural counterpart (39), and recent work (12) shows that linear controllability predictions are consistent with simulations of neuronal networks with Wilson-Cowan nonlinear dynamics.

Note that linearising neuronal dynamics along a limit cycle can lead to a time-varying Jacobian matrix **A**. In (4) we assume that the changes occur only in link weights and that the structure of matrix **A** is constant, being encoded by the *C. elegans* connectome. This allows us to apply *structural* controllability which is link-weight independent, hence variation in the nature or the strength of the links has no impact on our results as long as the network diagram remains unchanged. Ultimately, these modelling assumptions are simplifications motivated by theoretical tractability. Although the reasons above suggest that this framework might be usefully applied, it is the experimental validation that provides proof of this utility.

**What role do individual neuronal dynamics, and the resulting self-loops, play in controllability?**

Neurons have intrinsic dynamics. Activity is observed in the absence of external input, and response to external input is mediated by factors such as the neuron's own state, or membrane potential. In a network sense, such intrinsic dynamics manifest as self-loops on the nodes, as a self-loop represents a node's interaction with itself (see Figure 3). It has been claimed that any network where each node has a self-loop is structurally controllable (40). It is imperative to clarify that this result, namely that a single external signal can control the whole network with nodal self-loops, was derived under the specific condition that *the same single signal is directly imposed on every node*. In contrast, in the *C. elegans* nervous system only a small number of sensory neurons receive a given stimulus, hence not all nodes receive the same external signal. Given the two distinct components of the *C. elegans* connectome – neurons and muscle cells – we could use three distinct assumptions for self-loops: (i) All nodes have the same dynamics, and thus the same self-loops; (ii) Neurons have one type of self-loop and muscles have another; (iii) All nodes have different self-loops regardless of type. The first is the least realistic case, and the third the most. A recent paper (41) showed that if each node has an identical nonzero self-loop, as in (i), then controllability of a network can be estimated using the maximum matching framework derived in (1,4,2). This is what makes the calculation computationally tractable for networks with more than ~50 nodes. In (4), we proved that, for the target controllability of the *C. elegans* connectome, the assumption of identical self-loops for each node is not necessary and can be relaxed to case (ii), demonstrating that the results are valid even when assuming that different kinds of nodes (neurons or muscles) have different self-loops. The third and most fully realistic case (iii) where each neuron and each muscle exhibits different dynamics and therefore self-loops has yet to be explored, and represents a potential extension to the framework.

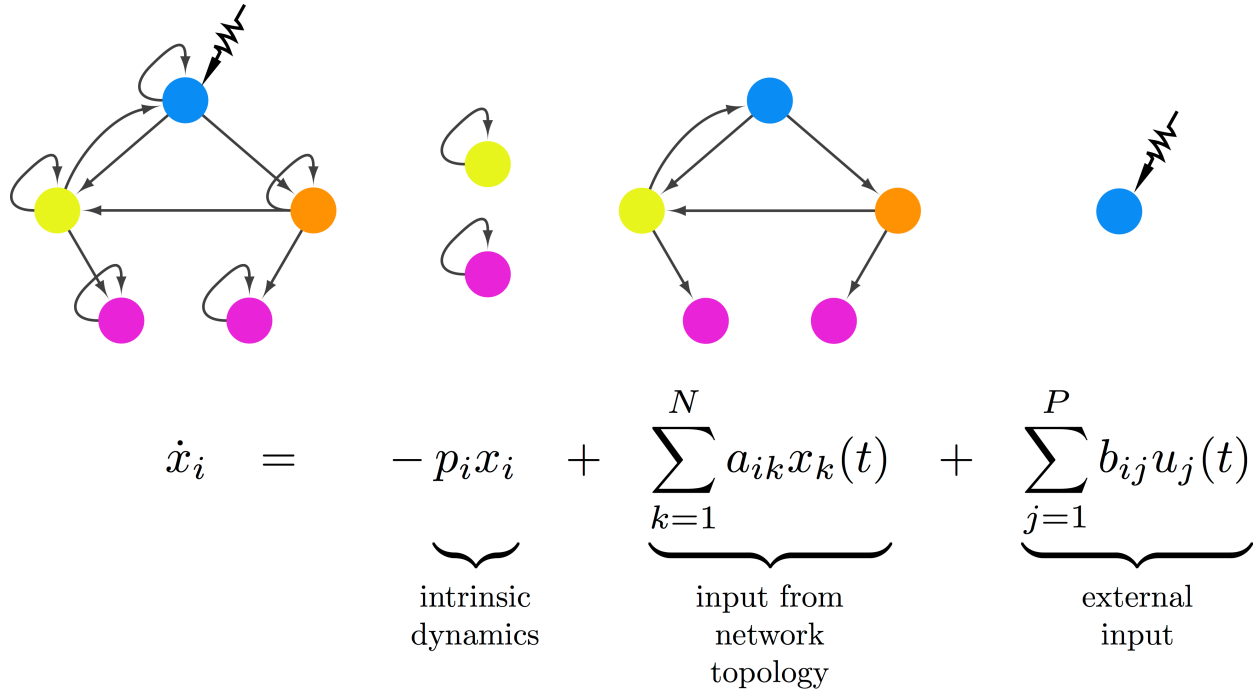

$$\dot{x}_i \quad = \quad \underbrace{-p_i x_i}_{\text{intrinsic dynamics}} + \underbrace{\sum_{k=1}^{N} a_{ik} x_k(t)}_{\text{input from network topology}} + \underbrace{\sum_{j=1}^{P} b_{ij} u_j(t)}_{\text{external input}}$$

**Figure 3: Intrinsic dynamics and self-loops.** The dynamics of many real networks – including neuronal networks – may be modelled as a simple set of ODEs (68,69), with terms to account for (i) the intrinsic dynamics of the nodes; (ii) the input signals from other nodes in the network resulting from network topology; and (iii) any external input signals. The intrinsic dynamics manifest as self-loops. We can assume that neurons have one type of self-loop, and muscles have another (4).

## COMPUTATIONAL AND EXPERIMENTAL DETAILS

**How are the analyses implemented computationally?**

We have made the code for the control analysis in (4) available in the form of Python scripts at https://github.com/EmmaTowlson/c-elegans-control. In its current form, it recovers the neuron classes predicted by the control analysis to be involved in the locomotory response to posterior and anterior gentle touch. Some simple amendments by the user will allow for the varying of input neurons. We are actively augmenting the repository to expand its capabilities.

**What are eigenworms and how can we use them to detect reductions in controllability?**

Previous studies have shown that the space of shapes adopted by *C. elegans* during motion is low dimensional, with just four dimensions accounting for 95% of the shape variance. This four-dimensional (42,43) eigenworm basis provides a compact, relatively unbiased representation of movement with which to look for phenotypes. However, it does not provide a direct measure of the movement dynamics of all 95 individual muscles. Indeed, abnormalities in some muscle groups may be easier to detect using the eigenworm basis than others. For example, the *unc-2* (neuronal voltage-gated $Ca^{2+}$ channel) which is strongly defective in neurotransmission from <u>all</u> neurons (see Figure 4, reproduced with permission from (44)) show very strong abnormalities in locomotion but usually overlap in their eigenworm statistics with wild-type worms. In the case of PDB, we observed a highly significant reduction in the ventral bias of omega turns, from 86% to 66% for PDB-ablated animals (4). Since the first eigen projection (EP1) represents the relative curvature of the entire body, the ventral bias in omega turns also translated into smaller, but measurable and statistically significant changes in EP1. This highlights the need for an in-depth screening of behavioural changes when assessing ablation phenotypes. In particular, it would be useful to develop better analytical tools to identify differences in shape dynamics at particular points along the worm body, to more directly correlate locomotion phenotypes with alterations in the activity of specific muscle groups.

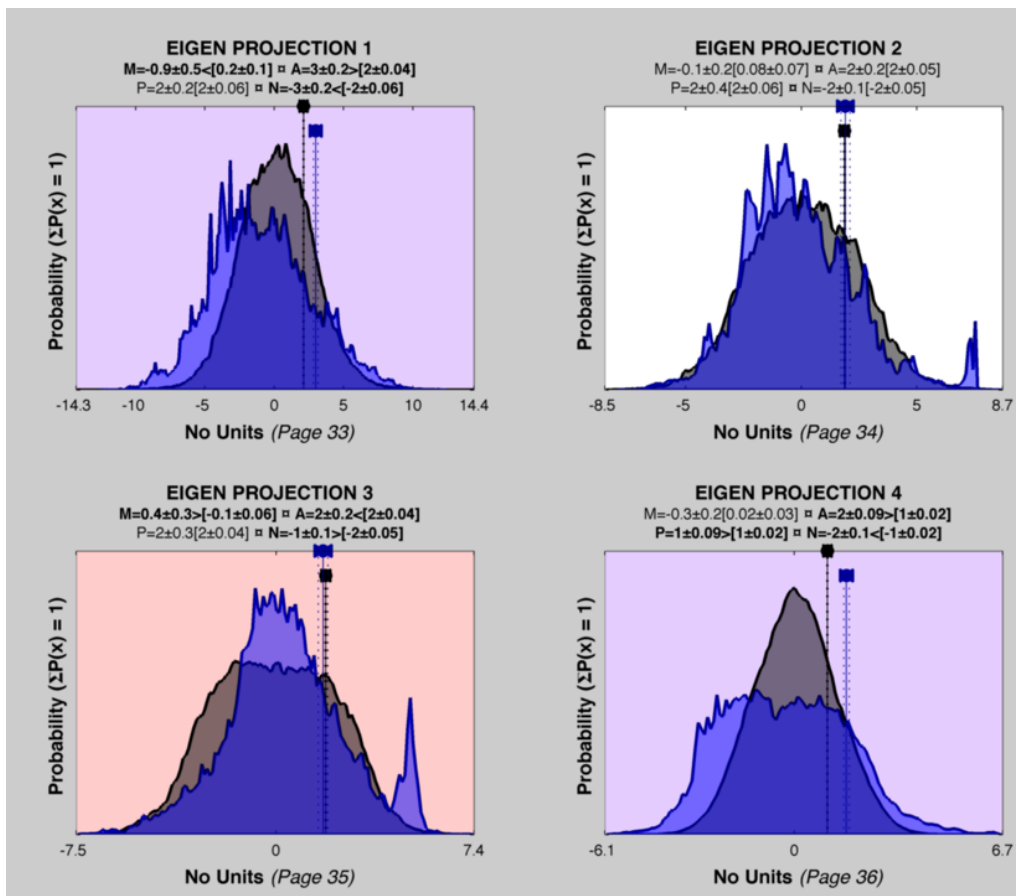

**Figure 4: Eigenworm feature statistics for the voltage-gated calcium channel mutant *unc-2*.** This phenotype is strongly defective in neurotransmission from *all* neurons, yet the eigenworm statistics overlap with wild-type worms. Reproduced with permission from the database described previously (44).

## POTENTIAL IMPROVEMENTS

**The connectivity data is likely inaccurate – how does it affect the analysis?**

The extraordinary mapping efforts by (21) have still not been rivalled 30 years later, and continue to fuel new insights on *C. elegans* (4,25). Remarkable as this very special dataset is, however, it is naturally imperfect. EM sections were reconstructed from five partial worms – primarily N2U and JSE (adult hermaphrodites), then N2T for the anterior nerve ring (adult hermaphrodite), N2Y (adult male) for the section between N2U and JSE, and finally JSH (L4 larva) to check connectivity in the nerve ring – see Figure 1. Whilst connectivity is widely assumed to be deterministic and almost invariant between individuals, it is highly unlikely to be precisely identical.

These issues – the compilation of worms, with not just individual variation, but different ages, and genders, plus the tracing errors inevitable with any mapping methodology – lead to a dataset that likely contains a number of erroneous and/or non-reproducible elements. Given the confounding factors, the level of unreliability is hard to quantify and can be expected to include missing connections, extra connections, mislabeled connections, and errors in the type and number of connections between neurons. Despite these issues, the existing connectivity map is clearly invaluable for an array of purposes and lines of inquiry (4,23,25). The accuracy of the control-based predictions offers another testament to its suitability. We tested the adequacy using a robustness analysis (4), finding that the predictions from the control framework are consistent and robust even when the data used for modelling contain discrepancies compared to the real connectome structure. Here, we extend the analysis presented in the original paper, finding that the predictions are robust to 420 random weak link deletions – see Figure 5. The fact that we can delete as many as 14% of the weak links, and rewire or add as many as 3% (4), and still recover our predictions, suggests that the worm employs a robust control framework – i.e. its wiring is such that it can maintain controllability even taking into account a high level of variation in the fine detail of the connectivity patterns.

A small section of the posterior worm body was never reconstructed. Beth Chen's thesis (2007) contains the most complete reconstruction of the ventral cord (where DD inter-neuronal synapses lie) as well as a parsimonious model of the dorsal cord (with the DD NMJs). Specifically, this work reconstructed many missing connections in the ventral cord, some via new EM images of thin sections from the original N2U worm. With respect to neuromuscular junctions, (to quote the source) "Neuron-to-muscle connections for the first 32 muscles in the head are detailed by (45). For the remaining muscles, direct neuron-to-muscle mapping is not available. In this case, we assume that motor neurons connect to muscles where positions of neuromuscular junctions overlap the sarcomere region of a given muscle… For neurons lacking complete reconstruction, especially ones on the dorsal side of the worm, the number of neuron-to-muscle connections is assumed to be the average NMJ per muscle from fully reconstructed neurons of the same class." However, there are still many synapses missing, prompting us to ask if this affects the controllability predictions. Most of the predictions in (4) stemmed from patterns of neuromuscular connections, where the estimates by (23) were based on the observed or interpolated positions of muscle arms and motorneuron processes, and thus are likely to be close to the real situation. However, it is certainly possible that the missing connectivity data has led to incomplete or even inaccurate predictions. Some efforts to infer this information from repeating patterns in the wider locomotor circuit have produced a probabilistic model describing the connections (46,47), which could be utilised as an estimate of connectivity in future analyses. In (4) we used the currently accepted gold standard dataset, but we do expect improved data to be made available over the next few years, including also more complete data filling this gap in the future.

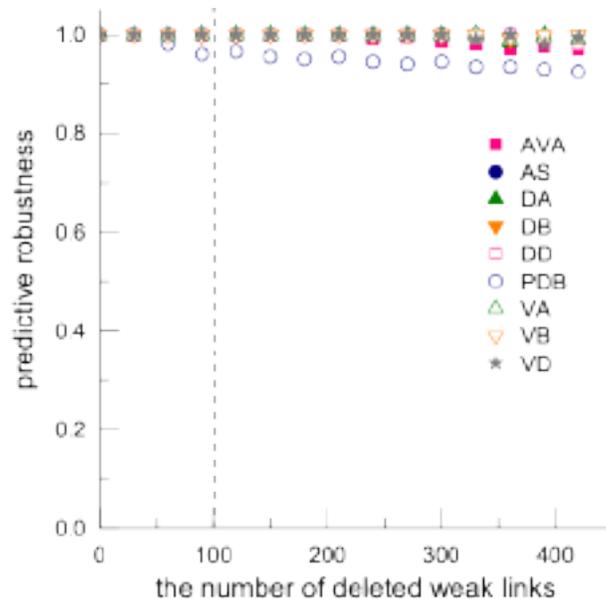

**Figure 5: Robustness of the predictions to imperfect data.** Robustness analyses, testing the robustness of ablation predictions to up to 420 random weak link deletions (~14% of the network).

**What happens to the control predictions if gap junctions are excluded from the network?**

Above, we noted that the chemical synapses are better modeled by the control framework than gap junctions are. To see how the inclusion of electrical synapses might have affected control predictions, we reanalysed the network using only the chemical synapses, and ignoring gap junctions. This makes the network sparser, which typically leads to more control nodes (1). Indeed, the analysis identified several additional neurons and neuronal classes predicted to affect controllability (see Table 1). Note that we recover all of our original predictions based on the aggregate network (in red), lending further credence to our original findings (4). We also identify three more neurons – AVB, PVC, and RID – when considering muscle control, all three of which have now been implicated in locomotion (48,49). Two of the four new neurons predicted for control of motor neurons have known roles in locomotion (49), and the final two, AVJ and PVR, have not currently been implicated and open an avenue for experimental testing in future work. Arguably, this larger set of predictions makes the synaptic only network more indicative of locomotion control and appropriate for the current set-up.

| Control | | Predicted Neuron Classes |
|---|---|---|
| *Control Muscles* | | DA, DB |
| | | DD |
| | | AVA, AVB |
| | | VA, VB, VD, AS |
| | | PDB |
| | | PVC |
| | | RID |
| *Control Neurons* | *Motor* | AVA, AVB |
| | | AVD |
| | | AVE |
| | | PVC |
| | | AVJ |
| | | DVA |
| | | PVR |

**Table 1:** Control predictions for the chemical synapse network; red font denotes the results for the aggregate network.

*FUTURE PERSPECTIVES*

**Is the control framework generalisable to other behaviours?**

The control framework is not limited to mechanosensation, it can be applied to investigate the basis of any behaviour that can be described as a response (output) to a stimulus (input), with known input and output nodes. For example, if we were interested in the locomotor response to olfaction, the output nodes would remain the muscles and we would define the input nodes as chemosensory neurons, such as ASH for the case of certain aversive chemicals (50). Note that while this mapping to the *target* control framework adequately models the majority of behaviours, control theoretic methodologies can be used without specifying the inputs and outputs (51,52).

An interesting extension of the current framework which may offer finer insights into different behaviours would be to take into account the time and energy required for specific control tasks (3,31,32,53), such as steering the muscles into a particular pattern of muscular activity. If the ablation of a component of the connectome leads to a significant increase in control time and/or energy for such an output state, then the component is likely involved in this behaviour. With the hypothesis that lower energy states that are reached within reasonable time are favoured over others, such temporal and energetic considerations could also elucidate why certain behavioural states are realised over others, and potentially offer insights into the extremely low dimensionality of the system (see above on eigenworms). To quantify the change in control time/energy we would need to include the weight and inhibitory/excitatory nature of each link and self-loop. Such information has not been available so far.

Finally, we note that a behaviour might involve different sets of sensory neurons, or different connections strengths at different times (for example in the case when connection weights are altered during learning). In this case, the problem could be reformatted using a *temporal* control framework, i.e. the input nodes, output nodes, and even the network structure itself may change with time. Recent efforts address the problem of *full* controllability of temporal networks (i.e. controlling all nodes in a temporal network) (54,55,32). The *target* controllability of temporal networks (controlling only the output nodes in a temporal network) remains an open but tractable theoretical question.

**Is the control framework generalisable to other organisms?**
The control framework is agnostic to the system in question, hence it is not limited to *C. elegans*. Currently, the main challenge presented by generalising the framework to other organisms is simply the availability, completeness, and accuracy of connectome data. Given that a full connectome at the neuronal level does not – for now – exist for many organisms (56) but partial maps and macroscopic maps do (13,15,16,57), we are left with two choices. The control framework can be applied at the meso- or macroscale resolutions, an approach that has already shown success (6). Alternatively, subcircuitry (58–60) – incomplete data in the sense of the whole brain, but complete when considered as a specific circuit – can be examined as smaller, standalone systems. For example, in the case of the *Drosophila* larva, we have knowledge of wiring of the olfactory glomeruli. Olfactory receptors (input) receive external stimuli in the form of odours, the information is processed by the local circuitry (control system), and descending neurons (output) pass signals to the mushroom body and elsewhere in the brain. Encouragingly, the previous robustness analyses indicate that in the face of imperfect or incomplete data, we can still glean important insights from the control framework.

**Discussion and outlook**
*C. elegans* remains at the forefront of the quest to understand the structure and function of brain networks, and offers an excellent model system from the perspective of control as well. We interrogated the theory and experimental practice for understanding control principles in the *C. elegans* connectome. We find that this framework recovers an impressive set of neurons or classes of neurons experimentally known to be important for locomotion; but it is not a complete list. A number of neurons with clear ablation phenotypes (for example, SMB and RIV (61)) were not predicted by the current implementation of the control approach. However, the neurons that *are* identified as affecting controllability are very likely to affect real behaviour, and in some cases (like PDB and the posterior DD neurons) these predictions will be unexpected from casual inspection. Classically, ablation experiments have been guided by obvious traits in the connectivity – RIV makes numerous synapses only to the ventral head muscles so was expected to have the effect on the ventral bias of turns which was observed (61). In (4) some of the predictions recovered by the control framework were much less intuitive – PDB, which we demonstrated to have a similar ablation phenotype to RIV, is located in the tail and, whilst it does have an asymmetric connectivity profile, its synapses are very small in number. Effects on the ventral bias of omega turns were therefore much more surprising. Similarly, whilst the DD neuron class was expected to play a role in forward and backward locomotion (49), the prediction from the control framework that the individual dorsal DD neurons would be more important than the anterior neurons was highly unexpected.

Importantly, the work in *C. elegans* gives us clues to approaches that are unlikely to be successful. For example, applying control theory to find a set of driver nodes amongst the full set of nodes in the nervous system (analogous to approaches in human brain data (7)) produced results that are more difficult to interpret biologically (52). It is only when formulating our question in terms of target control, and including the muscles as output nodes within our network that we began making successful predictions. This suggests also for higher order organisms that the emphasis should be on formulating control problems quite precisely, for example trying to steer network dynamics in a particular region (output nodes) away from particular aberrant dynamics (e.g. epileptic activity (11). We may also find that features and properties of other brain networks such as non-linearity and proprioceptive feedback become more important, and their incorporation will be an important challenge to add to the framework.

Over the next 5-10 years, theoretical and experimental advances will doubtless improve upon our current approach, and perhaps most significantly, so will new datasets. Neurotechnologies continue to be developed at an astonishing rate (14,62,63), and with them comes the promise of data which will complement, and ultimately supersede, the existing connectome. Functional data such as from calcium imaging may provide us with the weights and even signs on the connections needed to calculate *control energy* and *control time* (31,32,53), and investigate more mechanistic queries, such as how control is achieved, and why certain behavioural states are preferred over others. As larger and more detailed datasets arrive for higher order organisms, we will be equipped with the tools developed for *C. elegans* as the groundwork for elucidating their system-specific control principles.

**Keywords**
*C. elegans*; Network Science; Control theory; Locomotion; Connectome.

**Acknowledgments**
Funding for E.K.T. and A.-L.B was provided by the NSF under award number 1734821. P.E.V. was supported by the Medical Research Council grant number MR/K020706/1 and is a Fellow of MQ: Transforming Mental Health, grant number MQF17_24. G.Y. is supported by Thousand Young Talents Program of China. Y.LC. is supported by an EMBO Long-term fellowship, award number ALTF 403-2016. W.R.S. and D.S.W. are supported by Medical Research Council grant number MC-A023-5PB91 and Wellcome Trust Grant number WT103784MA.

**Authors' contributions**
A-L.B. and E.K.T. conceived of the manuscript. E.K.T., P.E.V., G.Y., Y.L.C, D.S.W., W.R.S., and A.-L.B wrote the manuscript. G.Y. wrote the code and E.K.T. edited it.

**Competing interests**
We have no competing interests.